\documentclass[12pt]{iopart}

\usepackage{graphicx}
\usepackage{hyperref}
\begin{document}

\title{Cryogenic surface ion trap based on intrinsic silicon}

\author{Michael Niedermayr$^1$, Kirill Lakhmanskiy$^1$, Muir Kumph$^1$, Stefan Partel$^2$, Johannes Edlinger$^2$, Michael Brownnutt$^{1}$ and Rainer Blatt$^{1,3}$}

\address{$^1$Institut f\"ur Experimentalphysik, Universit\"at Innsbruck, Technikerstr. 25, 6020 Innsbruck, Austria}
\address{$^2$Forschungszentrum Mikrotechnik, FH Vorarlberg, Hochschulstr. 1, 6850 Dornbirn, Austria}
\address{$^3$Institut f\"ur Quantenoptik und Quanteninformation, \"Osterreichische Akademie der Wissenschaften, Technikerstr. 21A, 6020 Innsbruck, Austria}
\ead{michael.brownnutt@uibk.ac.at}

\begin{abstract}
Trapped ions are pre-eminent candidates for building quantum information processors and quantum simulators. To scale such systems to more than a few tens of ions it is important to tackle the observed high ion-heating rates and create scalable trap structures which can be simply and reliably produced. Here, we report on cryogenically operated intrinsic-silicon ion traps which can be rapidly and easily fabricated using standard semiconductor technologies. Single $^{40}$Ca$^+$ ions have been trapped and used to characterize the trap operation. Long ion lifetimes were observed with the traps exhibiting heating rates as low as $\dot{\bar{n}}=$ 0.33\,phonons/s at an ion-electrode distance of 230\,$\mu$m. These results open many new avenues to arrays of micro-fabricated ion traps.
\end{abstract}

\maketitle

\section{Introduction}
Processing information using a quantum system allows problems to be solved which are difficult or impossible to compute classically \cite{Chuang}, and
trapped atomic ions provide a promising platform for such quantum information processors \cite{Ladd,Blatt,Zoller}.
Ions can be confined using a combination of radio-frequency (RF) and static (DC)
electric fields \cite{Werth}. The electrode structures required to create the appropriate fields can take a variety of geometries, and great progress has recently been made using planar structures \cite{Chiaverini,Hughes2011}.
For these, all electrodes are located in a plane above which the ions are trapped. Notable among the advantages of such a structure is the scalability \cite{Kim}: arrays with hundreds or thousands of traps can in principle be built on a single chip using very-large-scale integration (VLSI) technologies like optical lithography and etching, far exceeding what is possible using bulk fabrication.

In making such traps a reality, a number of features must be considered. It is essential that scalable trap architectures have low RF losses: if the losses are too high the RF power applied in order to trap will be dissipated in the chip. Additionally, microfabrication has become highly developed and it would be desirable to bring these techniques to bear on ion traps. For large arrays of traps the creation of through-wafer vias \cite{Motoyoshi}, so that electrical connections can be made throughout the array, is a prerequisite. Optical addressing of ions in an array can be facilitated by fabrication of holes through the substrate. Moreover, the integration of on-chip electronics, such as would be afforded by CMOS (Complementary Metal-Oxide-Semiconductors), would open up untold possibilities for ion traps \cite{Mehta14, Alonso13}.

Surface ion traps can be broadly divided into two groups, according to the
type of substrate material: dielectric or semiconductor. Dielectric substrates like sapphire or fused silica have a very low RF power dissipation. For such traps the electrodes are created by simple optical lithography combined with lift-off processes or electroplating on top of the substrate \cite{Seidelin}. However, dielectric materials are difficult to pattern by means of wet or dry etching and not suitable for features such as through-wafer vias. The second type of substrate materials are semiconductors, such as silicon, which are easy to structure. Slots, holes and vias with aspect ratios up to 160 can be etched by standard techniques \cite{Parasuraman}. However, the RF losses in intrinsic (i.e. undoped) silicon at room temperature are is significant.

In some cases, the issue of RF losses was mitigated by using highly doped
silicon for the trap electrodes \cite{Sterling, Britton} or substrate \cite{Wilpers}. This method would not, however, work in a cryogenic environment because of the silicon's low electrical conductivity at low temperatures. In other cases there was an additional ground electrode which shielded the silicon against the trapping RF voltage \cite{Allcock, Doret, Leibrandt09}. This type of trap can also be operated at cryogenic temperatures \cite{Vittorini13}, though it necessitates a more complicated fabrication process and precludes vias for RF electrodes.

At low temperatures, the charge carriers in intrinsic silicon freeze out,
leaving the substrate as a good insulator with low RF loss. This obviates
the need for a shielding electrode. By operating intrinsic silicon traps at
cryogenic temperatures and omitting the shielding ground-plane, the trap capacitances and the amount of power dissipated can be reduced. A range of
fabrication techniques also become available, which would have otherwise
have been precluded either by the substrate material or the ground plane.
Intrinsic silicon is already used for several superconducting-qubit
applications in the mK-range \cite{Chang, Johnson12}, though all silicon-based ion traps to date have been designed to operate at room-temperature.

In this paper, we report on silicon-based surface ion traps built for
cryogenic applications. The cryogenic environment inheres several advantages
which are well known. Ultra-high vacuum can be attained within a few hours
due to cryogenic pumping, without baking the system \cite{Antohi}. This facilitates fast turn-around times for trap installation ($\sim$1 day). Furthermore, operation at liquid-helium temperatures reduces the rate at which the ions' motion is heated, typically by around two orders of magnitude \cite{Labaziewicz}. This is beneficial as it increases the coherence time of the ions' motion, which is used to transfer quantum information between different ions. In addition to these benefits, the novel use of intrinsic silicon as a trap material means that the traps neither need nor have a ground plane to shield the substrate. This allows for a very simple basic fabrication procedure and permits the use of standard silicon processing techniques, while mitigating the problems of RF loss.

\section{Design overview and fabrication}
\begin{figure}
\centerline{\includegraphics[width=10cm]{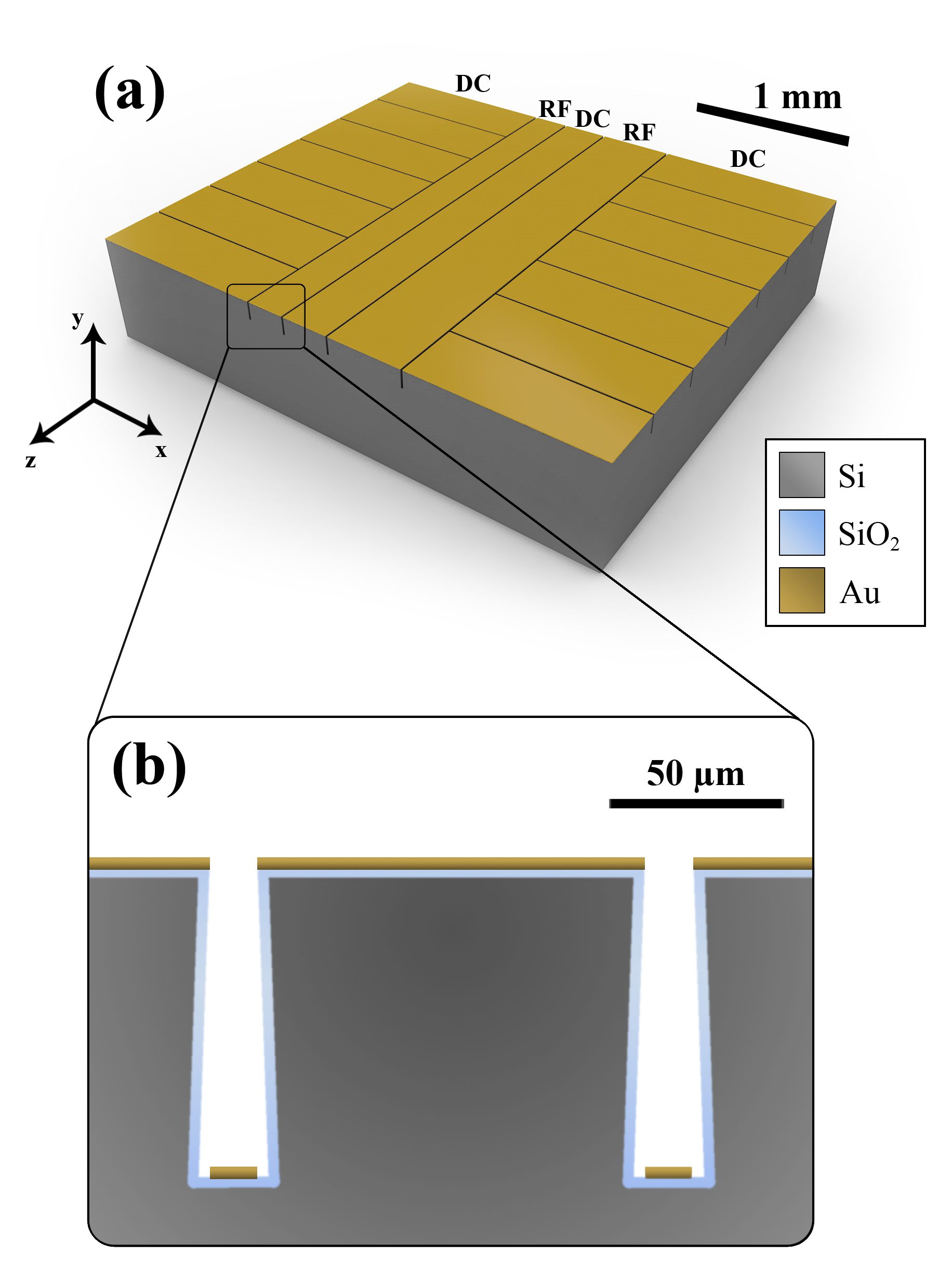}}
\caption{(a) Schematic showing the surface-trap design. The trap consists of a pair of asymmetric RF electrodes (widths of 200\,$\mu$m and 400\,$\mu$m respectively), a centre DC electrode (width: 250\,$\mu$m) and  seven segmented DC electrodes on each side (axial width: 350\,$\mu$m). The electrode separation is 10\,$\mu$m. This geometry allows strings of ions to be trapped 230\,$\mu$m above the surface. (b)~Cross section through the trap. Trenches etched to a depth of $\sim$100\,$\mu$m separate the individual electrodes. The entire silicon surface is covered by a thermally grown SiO$_2$ layer preventing metals from diffusing into the silicon. The trap electrodes are created by gold deposition normal to the surface. Well-defined undercuts prevent electrical connection between the different electrodes.}
\label{fig:figure1}
\end{figure}

The planar design used for this work is illustrated in figure\,\ref{fig:figure1}. The electrode layout is based on a similar design used elsewhere \cite{Daniilidis2}. Ions are trapped 230\,$\mu$m above the centre electrode by applying an RF voltage to the two RF electrodes. Lasers used to Doppler cool the ions are aligned parallel to the plane of the trap, to minimize scatter from the surface. To efficiently cool the ions along all three principal axes of their motion, each axis must have a projection along the direction of the cooling laser beam. Consequently, none of the principal axes of motion should be perpendicular to the trap surface. The RF electrodes have thus been fabricated with unequal widths ensuring that the two radial principal axes both have a component parallel to the trap surface \cite{Shaikh}. Seven segmented electrodes are located on each side of the two RF electrodes. DC voltages are applied to these electrodes and create confinement along the axial (\textit{z}) direction. Ions can be shuttled along the \textit{z}-axis by changing the DC voltages. 
Stray electric fields arising from contamination of the trap or imperfection in trap fabrication can cause micromotion \cite{Berkeland}, which is minimized by adding suitable DC voltages to the segmented DC electrodes.

\begin{figure}
\centerline{\includegraphics[width=15cm]{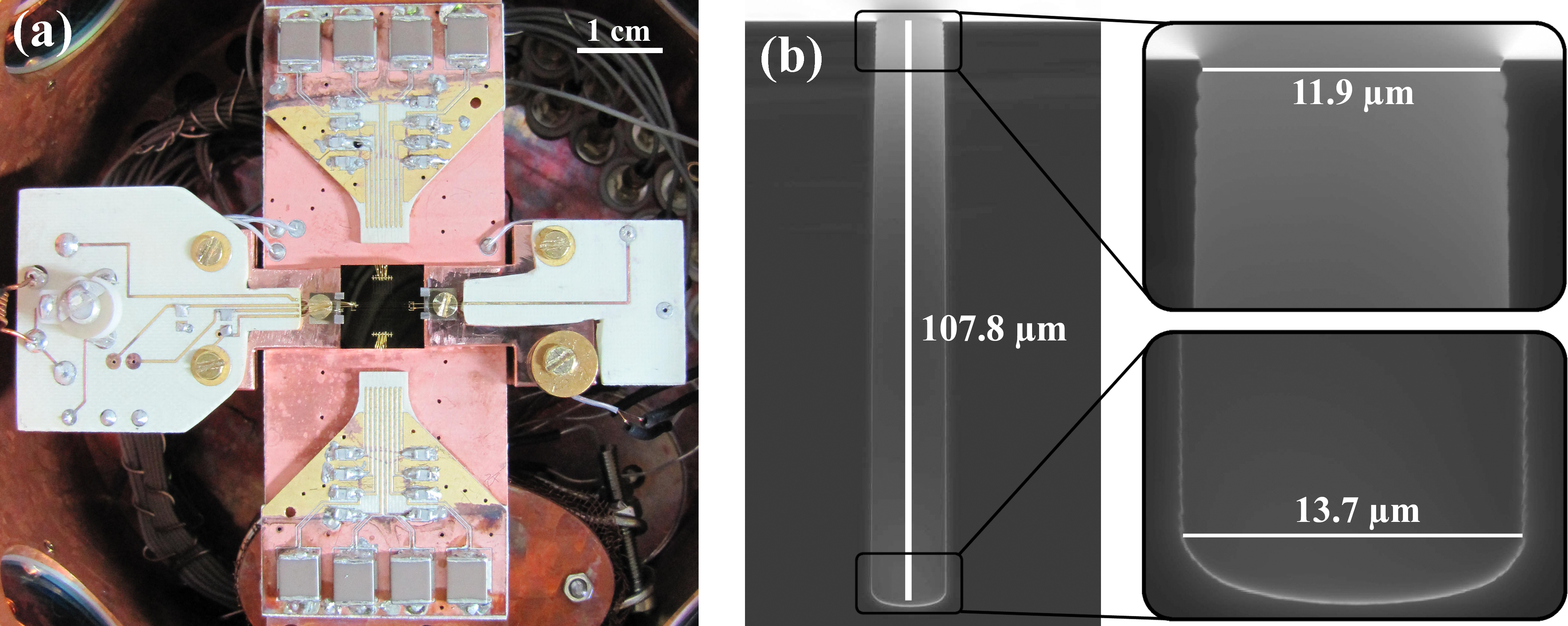}}
\caption{(a) Cryostat's 10\,K stage. The trap is at the centre. The DC filters are located on the vertical printed circuit board. The LC drive circuit is mounted on the left of the trap and a capacitive divider for measuring the trap voltage is on the right. (b) Scanning electron microscope images of an undercut trench formed by deep reactive ion etching separating two electrodes.}
\label{fig:figure2}
\end{figure}

To fabricate the traps an intrinsic silicon substrate is patterned by optical lithography and deep reactive ion etching (see figure\,\ref{fig:figure2}(b)). Deep, undercut trenches are etched in the substrate and mark the gaps between the individual electrodes. Following a thermal growth of 2\,$\mu$m of SiO$_2$ the electrodes are formed by evaporation of gold perpendicular to the surface.
Due to the undercuts, there is no electrical connection between the different electrodes (figures\,\ref{fig:figure1}(b), \ref{fig:figure2}(b)) and no further lift-off or etching steps are necessary. There are no further cleaning steps and, to avoid any contamination of the surface, the gold electrodes are never brought into contact with liquids such as solvents for cleaning. The fabrication process and the packaging are explained in detail in \ref{app:app1}.

Initially, the operation of a single trap (trap \#1) was characterized in detail over several weeks. To demonstrate the reproducibility of the results, a further five traps - randomly selected from two different silicon wafers - were tested and shown to exhibit similar behaviour.

\section{Cryogenic setup}
The silicon ion trap is cooled to 10\,K in a closed-cycle, two-stage, Gifford McMahon cryostat equipped with a vibration-isolation system \cite{Antohi,Gandolfi} to reduce the vibrations at the trap to around 100\,nm at 2\,Hz.
The trap is attached to the second cooling stage of the cryostat (figure\,\ref{fig:figure2}(a)). It is enclosed by a copper shield, also cooled to 10\,K. This minimizes the incident black-body radiation and reduces the number of background molecules at the trapping site, as they freeze on the shield walls. 
Surface contamination caused by background gas molecules freezing out on the trap electrodes during the cool down can be reduced by temporarily heating the trap. Since the background gas mainly consists of water, it is sufficient to keep the trap at 320\,K till the first stage of the cryostat has reached a temperature of 240\,K. 
Thereafter the second stage is cooled to 10\,K, while the first stage reaches a final temperature of 50\,K. $^{40}$Ca$^+$ ions are loaded from a neutral Ca-beam produced by a resistively heated oven located within the vacuum chamber but not mechanically connected to the cold stage. The atoms are introduced to the trapping region through a small hole in the copper shield (diameter $\sim$3\,mm) and are ionized by a two-photon process \cite{Gulde} in the trapping region.

The voltage on the RF electrodes (amplitude $U_0 = 140$\,V, frequency $\Omega_\mathrm{T}/2\pi = 20.6$\,MHz) is provided by an LC lumped-circuit resonator driven by a function generator \cite{Gandolfi} and creates a trapping potential with trap depth of 75\,meV. 
The power dissipation in the resonator goes as $P_\mathrm{D} = U^2_0C\Omega_\mathrm{T}/2Q$, where $C$ is the resonator capacitance and $Q$ the quality factor of the resonator. To keep $P_\mathrm{D}$ low, $C$ must be kept small. This is achieved by locating the resonator in vacuum next to the trap on the cold stage of the cryostat (see figure\,\ref{fig:figure2}(a)). The measured $C$ is 9.5\,pF, which is mainly limited by the capacitance of the RF electrodes. The power necessary for trapping is less than 10\,mW, which is well below the cryostat's cooling power of 500\,mW and increases the temperature measured next to the trap by only 0.2\,K.
Filters with a cut-off frequency of 4.8\,kHz are mounted on the cold stage next to the resonator in order to filter the voltages applied to the DC electrodes (see \ref{app:app2} for further details on the LC resonator and DC filters).
DC voltages in the range -40\,V$<V_\mathrm{DC}<$40\,V are applied to the central three electrode segments and provide an axial trapping frequency of around 1\,MHz. With this voltage configuration the principal radial axes are tilted with respect to the $x$ and $y$ directions by approximately 20$^{\circ}$.

\section{Silicon at cryogenic temperatures}
\begin{figure}
\centerline{\includegraphics[width=12cm]{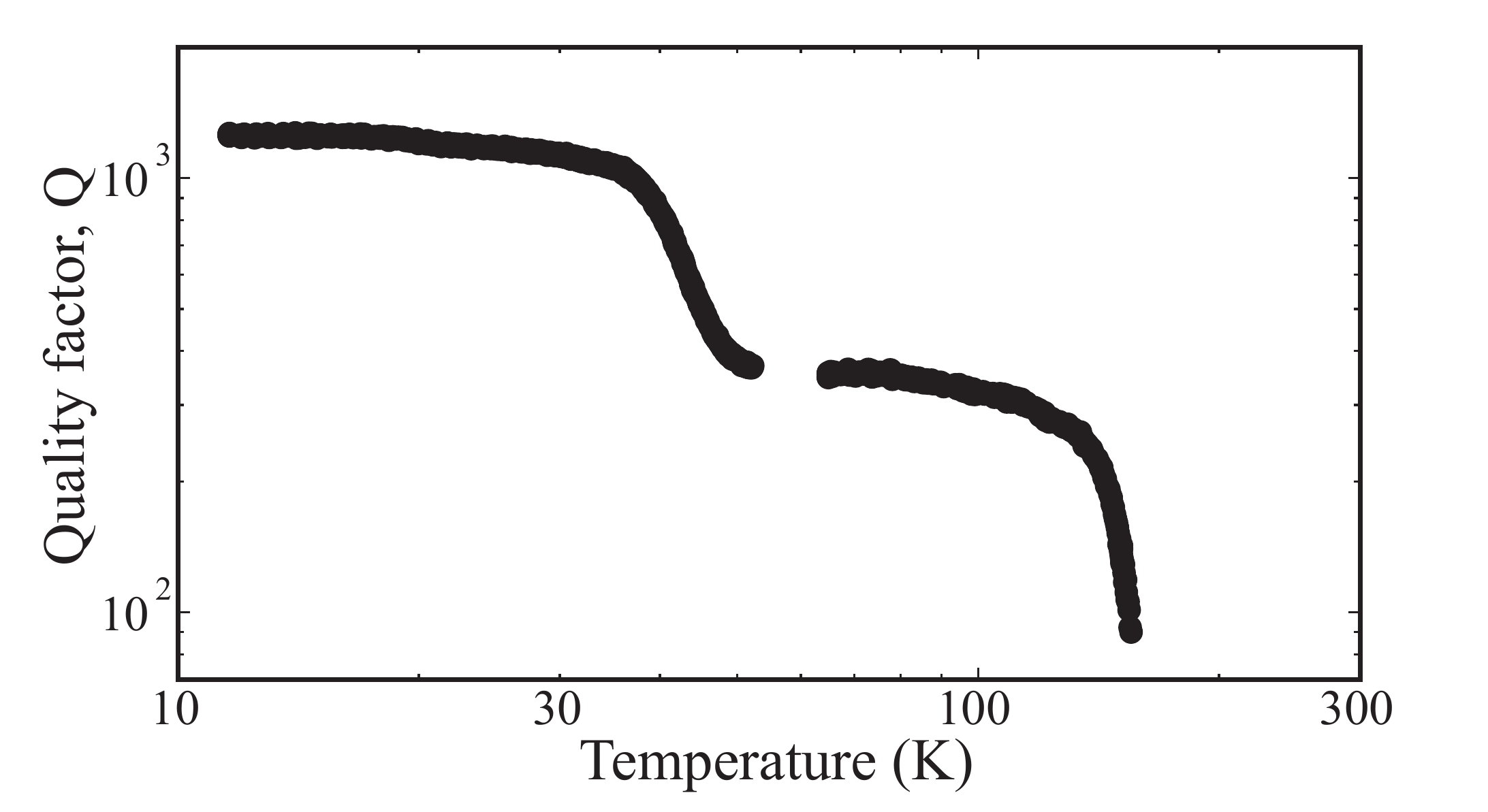}}
\caption{Resonator quality factor as a function of temperature. The inductive and capacitive parts of the LC resonator are predominantly provided by a copper air-coil and the trap RF electrodes, respectively. At room temperature, the entire RF driving power is absorbed by the silicon substrate and there is no measurable resonance. On cooling, the conductivity of the inductor increases (thereby increasing its quality factor, $Q_L$). Cooling also reduces the charge-carrier concentration in the silicon and, below 100\,K, causes a steep decrease of its electrical conductivity and loss tangent (thereby increasing the capacitive quality factor, $Q_C$). These effects all serve to increase the overall $Q$, with the plateau around 100\,K being due to the non-linear response of the material properties. Below 20\,K, the quality factor is comparable to that measured with a fused-silica trap, meaning that $Q$ is limited by $Q_L$ and that $Q_C \gg Q_L$. The data shown were measured in trap \#1 (cf. table\,\ref{tab:table1}). The other five traps tested showed similar behaviour.}
\label{fig:figure3}
\end{figure}

The characterization of the RF resonator (including the trap) as a function of temperature is shown in figure\,\ref{fig:figure3}. The inductor of the LC resonator is provided by a copper coil with an air (vacuum) core and the capacitance primarily comes from the trap RF electrodes. The quality factor, Q, was recorded while the cryostat was slowly heated up. The temperature was measured by a silicon diode mounted on the copper trap carrier. $Q$ follows the relation $1/Q=1/Q_\mathrm{L}+1/Q_C$, where $Q_\mathrm{L}$ and $Q_C$ are the quality factors of the inductor and the capacitor, respectively. 

At room temperature the silicon substrate, which supports the RF electrodes, has a very high loss tangent \cite{Krupka}, tan\,$\delta$, of 1.5 at the driving frequency $\Omega_\mathrm{T}/2\pi=$ 20.6\,MHz. For comparison, under the same conditions, tan\,$\delta$ of fused silica is around $\sim$10$^{-4}$ \cite{Kaye}.
Due to the high loss tangent, at room temperature the entire RF driving power is absorbed by the silicon substrate. There is no measurable resonance: using an impedance analyzer the measured resonator $Q$ was less than 20, and within the measurement uncertainty it was indistinguishable from zero. In contrast, a quality factor of 400 was measured in a similar trap fabricated on a fused-silica substrate and operated at room temperature. 
Cooling leads to a reduction of the charge-carrier concentration in the silicon and, below 100\,K, to a steep decrease of the electrical conductivity and loss tangent. Ultimately, all free charge carriers freeze out at $\sim$25\,K and the silicon becomes an insulator \cite{Krupka}. In addition to these changes in the silicon, the electrical conductivity of the coil increases with decreasing temperature, and therefore the inductor quality factor, $Q_\mathrm{L}$, goes up \cite{Gandolfi}. Increasing $Q_C$ and $Q_\mathrm{L}$ leads to an increasing overall resonator quality factor, $Q$, with decreasing temperature, as shown in figure\,\ref{fig:figure3}. Below 20\,K the value of $Q>1200$ is comparable to that measured with a fused-silica trap at the same temperature indicating that $Q$ is then only limited by $Q_\mathrm{L}$ and not by RF absorption in the silicon.

Photocharging of trap structures has previously been observed in a number of experiments \cite{Harlander2010, Wang2011}. In the traps reported here, it is in principle possible that photons could also excite charge carriers in the silicon of the trenches' sidewalls which are not covered by gold. Such an effect could disturb trap operation. To investigate this possibility, the resonator's quality factor was measured at low temperature both with trapping lasers on and off to determine the influence of the laser light. The trap was directly illuminated with around 10\,mW/cm$^2$ of laser light at each of the wavelengths used to trap $^{40}$Ca$^+$ ions (375\,nm, 397\,nm, 422\,nm, 729\,nm, 854\,nm and 866\,nm). However, no changes in performance were observed. Furthermore, while trapping $^{40}$Ca$^+$ ions, no effects due to photo-charging of the substrate were observed.

\section{Trapped ion lifetime}
A trapping parameter which becomes especially important with a large number of trapped ions is the length of time for which an ion can be trapped, sometimes called the trapped-ion lifetime. The loss of a single ion can complicate the implementation of, for example, a quantum algorithm. While trapping lifetimes of a few minutes may be acceptable for experiments using small numbers of ions, experiments with many ions require significantly longer ion lifetimes, as the probability of losing an ion increases. Ions are usually lost by collisions with background gas molecules or by motional heating. A cryogenic setup is an ideal tool to achieve long lifetimes since it provides extremely high vacuum and helps to reduce heating rates \cite{Antohi,Labaziewicz,Berkeland98}.

Cooled and uncooled ion lifetimes were investigated using trap \#1. With laser cooling, no ion losses were recorded over a total experimental period of more than 50\,hours with a single ion. 
To investigate uncooled lifetimes, the lasers were turned off and, after some waiting time, turned back on to see if the ion was still trapped. Waiting times of up to 9~hours were used, and the ions were never lost. 
The trap is therefore suitable for scaling up to hundreds of ions without the need for continuous reloading. Five further traps were tested for shorter periods and the results from these traps were consistent with the more extended observations made with trap \#1.

\section{Heating rates}
\begin{figure}
\centerline{\includegraphics[width=12cm]{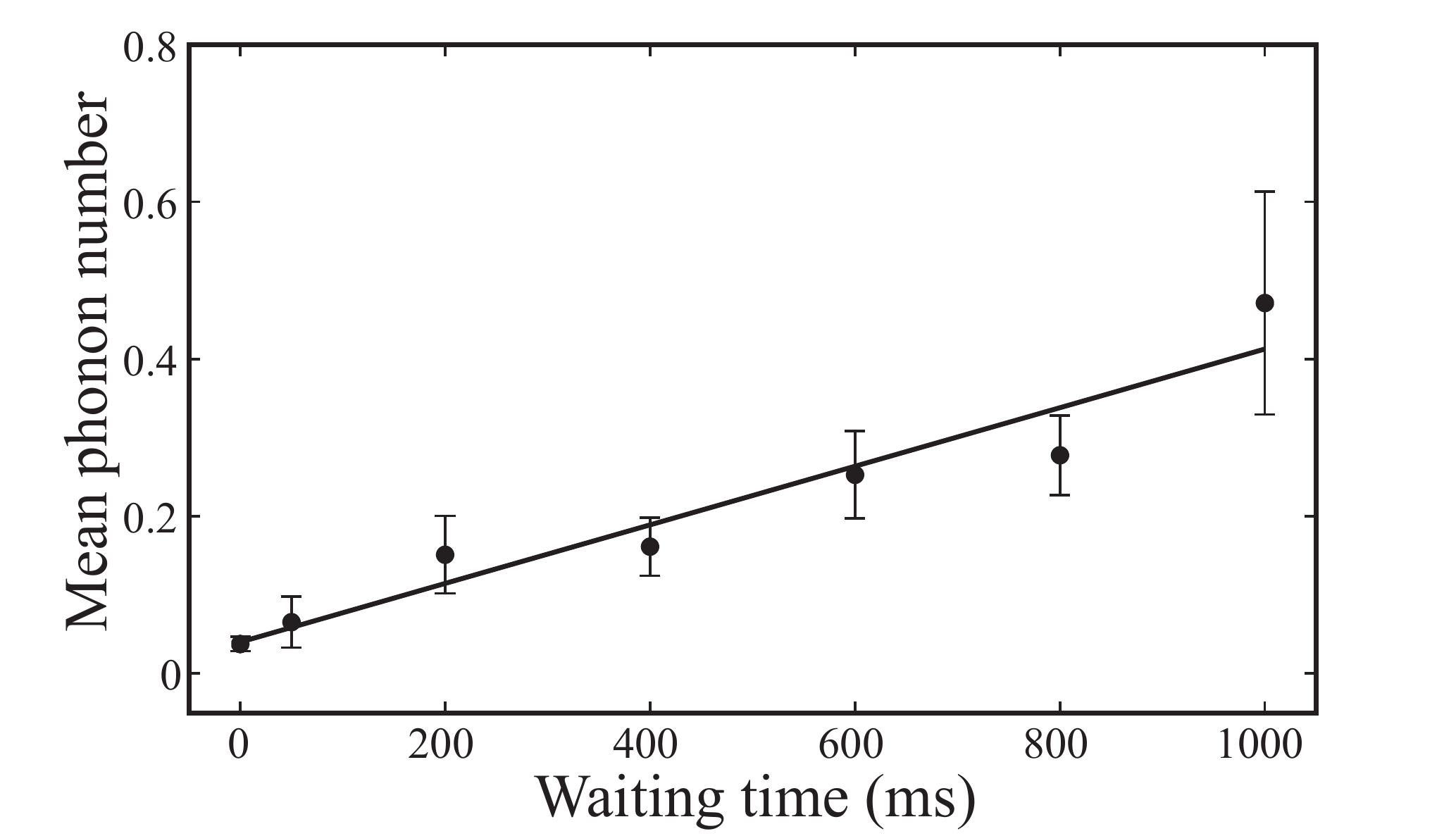}}
\caption{Heating-rate measurement. Mean phonon number $\bar{n}$ of the axial mode ($\omega_z/2\pi=1.069$\,MHz) as a function of the waiting time after ground-state cooling. $\bar{n}$ was determined by measuring the Rabi flops on the blue sideband \cite{Leibfried}. In this instance the heating rate, taken to be the gradient of a linear fit to the data, is $\dot{\bar{n}}=0.37(6)$\,phonons/s. Taking data on different days over a period of six weeks the trap exhibited a heating rate of $\dot{\bar{n}}=$ 0.6(2)\,phonons/s.}
\label{fig:figure4}
\end{figure}

For many quantum-information applications ions should be at or near their motional ground state, and heating of the ions' motion degrades the quality of quantum operations. Trapped ions are predominately heated by electric-field noise resonant with the ions' motional frequencies. 
To measure the heating rate \cite{Leibfried}, the axial motion of a single ion was cooled to near the ground state by resolved-sideband cooling. Following a predefined waiting time, the mean phonon number was determined by two different methods: measurement of the transition probability on the red and blue sidebands and Rabi flops on the blue sideband \cite{Leibfried}.
The heating rate was determined by the change in phonon number with different waiting times (figure\,\ref{fig:figure4}).

Over a period of six weeks, the heating rate in trap \#1 was measured several times and found to be constant, within the error bars, at $\dot{\bar{n}}=$ 0.6(2)\,phonons/s. The electric-field noise inferred to underlie this heating \cite{Labaziewicz} is $S_\mathrm{E}=4.4 \times 10^{-15}$\,V$^2$m$^{-2}$Hz$^{-1}$.
Heating rates were measured in five further traps, four from the same wafer and one (trap \#6) from a second wafer, with the results given in table\,\ref{tab:table1}.

\begin{table}
\caption{\label{tab:table1}Heating rates of different Si traps. Heating rates of six traps, all of the same design, with an ion-electrode separation of 230\,$\mu$m. Measurements were made over 84 days. Traps \#1-5 were taken from a single Si wafer, with trap \#6 being from a second wafer. 
The heating rate measured in trap \#5 is 0.33(4)\,phonons/s, which is the lowest rate ever reported.}
\begin{indented}
\lineup
\item[]\begin{tabular}{@{}*{3}{l}}
\br                              
Trap \# & Heating rate (phonons/s) & Axial freq. (MHz)\cr 
\mr
 \0\01 & \0\0\00.6(2)& \0\01.069\cr
 \0\02 & \0\0\03.3(2)& \0\01.059\cr
 \0\03 & \0\0\00.96(7)& \0\01.069\cr
 \0\04 & \0\0\00.95(7)& \0\01.045\cr
 \0\05 & \0\0\00.33(4)& \0\01.066\cr
 \0\06 & \0\021.5(8)& \0\01.073\cr
\br
\end{tabular}
\end{indented}
\end{table}

Trap \#5 exhibits a heating rate of 0.33(4)\,phonons/s, which is the lowest measured heating rate ever reported: The lowest previously-reported heating rate at room temperature was 0.83(10)\,phonons/s in a trap with a 3.5\,mm ion-electrode separation \cite{Poulsen}. The lowest previously-reported heating rate at cryogenic temperatures was 2.1(3)\,phonons/s in a trap with a 100\,$\mu$m ion-electrode separation \cite{Wang}. 
It is not known to what extent the lower heating rate in the present trap can be attributed to the use of silicon, rather than the lower temperature (compared to \cite{Poulsen}) or larger trap dimensions (compared to \cite{Wang}).
The reason for the variation in heating rates between traps on the same wafer is not known, though it is not necessarily surprising, given the level of variation which is often observed in nominally identical traps \cite{Labaziewicz}.  
Trap \#6 exhibits a heating rate of 21.5(8)\,phonons/s. This falls well within the range of heating rates observed in other cryogenic ion traps \cite{Labaziewicz,Labaziewicz2}, but is is significantly higher than the other five traps tested here. The second wafer, from which trap \#6 was taken, was nominally identical to the first, but was patterned separately. Slightly modified etch parameters were used, so that the underetch initially proceeded at a steeper angle. It is not known whether this is related to the higher heating rate.

\section{Conclusion and Outlook}
We have presented a new ion-trap design based on a silicon substrate. This will allow trap fabrication to benefit from well-developed silicon-fabrication processes. The design was made possible by exploiting the fact that, at low temperatures, intrinsic silicon becomes an insulator with low RF losses. The traps exhibit a high $Q$ of $>$1200 and reproducibly low ion-heating rates of around 1\,phonon per second at a trap frequency of 1\,MHz.

Fabrication could be extended to include slots for increased optical access \cite{Allcock} and through-wafer vias \cite{Motoyoshi}. Unlike room-temperature silicon traps, vias in our design would also work at RF frequencies, allowing the realization of a 2D trap array with adjustable RF electrodes \cite{Kumph}. One may ultimately envision such traps integrated with a wide variety of other silicon-based technologies including CMOS electronics \cite{Mehta14, Alonso13}, micro-optics \cite{Streed2011}, micro- and nano-mechanical systems, and sensors. This would ultimately mean that the entire science package, including optical, mechanical, and electronic functions could be integrated on a single substrate to provide a quantum lab on a chip.
 
\section*{Acknowledgements}
M.N. thanks M. Brandl for discussions about DC filters. S.P. thanks P. Choleva and S. Stroj for assisting during the trap fabrication. The authors acknowledge support from the European Research Council through the project CRYTERION \#227959 and the Institute for Quantum Information GmbH.

\clearpage

\appendix
\section{}

\subsection{Trap fabrication and packaging}\label{app:app1}
High-purity float-zone silicon wafers (diameter: 100\,mm, thickness: 525\,$\mu$m) with specific resistivity larger than 5000\,$\Omega$cm were used. The wafers were coated by the positive photoresist AZ1518 with a thickness of 2.4\,$\mu$m. The resist was patterned by means of standard optical lithography. The wafers were deep reactive ion etched by gas chopping based on SF$_6$ and C$_4$F$_8$ to create trenches with slight undercuts separating the different electrodes of the ion traps. The 10\,$\mu$m gaps between the electrodes were etched to a depth of $\sim$100\,$\mu$m with an undercut of $\sim$1\,$\mu$m. The resist was then removed by O$_2$ plasma cleaning. A 2\,$\mu$m thick SiO$_2$ layer was grown on the silicon surface by thermal oxidation to prevent metals from diffusing into the silicon. Each wafer provided 52 traps and the individual trap chips were separated by laser scribing. To form the electrodes, titanium and gold layers of a thickness of 2\,nm and 500\,nm respectively were deposited on the substrate surface by electron-beam evaporation. No further cleaning steps were performed after evaporation.

Traps were mounted on a copper carrier. To ensure good thermal contact between the trap and the carrier, each trap was coated on the backside with a thin layer of heat-conducting grease (Apiezon-N) and then clamped in place by two stainless-steel forks. Printed circuit boards (PCBs) (Rogers - RO4350B) supporting the DC filters and the LC resonator where glued (Stycast - 2850-FT) to the carrier. The trap electrodes were connected to the PCBs by 25\,$\mu$m thick gold wirebonds. The copper traces on half of each PCB were gold-electroplated to increase the adhesion of the bonding wires from the trap to the PCBs (see figure\,\ref{fig:figure2}(a)). The wiring of the cryostat to connected the PCBs to the outside world is explained in detail in \cite{Gandolfi}. The entire fabrication and assembly process of the trap took place in a cleanroom to reduce surface contamination. The only exception to this was installing the trap in the cryostat itself which was not located in a cleanroom, though this step took less than ten minutes.

\subsection{LC resonator and DC filter}\label{app:app2}
The lumped-circuit LC resonator is similar to that reported by Gandolfi et al.\cite{Gandolfi} and is formed by a homebuilt copper air-coil inductor with an inductance of 6.3\,$\mu$H mounted next to the trap, a capacitance of 9.5\,pF provided by the trap's RF electrodes, and a capacitive voltage divider. The circuit's resonance frequency is 20.6\,MHz at 10\,K. The capacitive divider, which allows the measurement of the voltage on the RF electrodes, has a ratio of 1:400 and a total capacitance of 2.5\,pF. It consists of one 1000\,pF capacitor (CDE - MC22FD102J-F) and two 5\,pF capacitors (CDE - MC12CD050D-F) arranged in parallel. Furthermore, there is a matching network to match the LC circuit to 50\,$\Omega$. This is located next to the resonator at the 10\,K stage. It consists of a tunable capacitor (12-100\,pF, Johanson Manufacturing - 9328) and a homebuilt inductor (186\,nH) connected in series and parallel, respectively.

For best trapping performance all DC electrodes must be properly RF-grounded which can be accomplished by installing capacitors as close as possible to the DC electrodes. For this reason, small surface-mounted NP0 capacitors with a capacitance of 470\,pF (Kemet - C0805C471J1GACTU) are located $\sim$15\,mm from the electrodes. Additionally, RC low-pass filters are used to filter RF noise on the DC lines. The filters each consist of a 100\,$\Omega$ thin-film resistor (VPG - Y1625100R000Q9R)
and a 330\,nF NP0 capacitor (Kemet - C2220C334J1GACTU) placed $\sim$30\,mm from the trap. The cut-off frequency of these filters is 4.8\,kHz. The resistors and capacitors used are cryo-compatible and do not significantly change performance during cooling. In addition to these filters there are 6th-order RC low-pass filters outside of the vacuum chamber with a cut-off frequency of 80\,Hz.

\section*{References}
\bibliographystyle{unsrt}

\end{document}